\documentclass{elsarticle}

\usepackage{amssymb}
\journal{Journal of High Energy Astrophysics}

\begin{document}

\begin{frontmatter}



\title{Numerical Simulations of Gamma-Ray Burst Explosions}


\author[label1]{Davide Lazzati}
\author[label2]{Brian J. Morsony}
\author[label3]{Diego L\'opez-C\'amara}

\address[label1]{Department of Physics, Oregon State University, 301
  Weniger Hall, Corvallis, OR 97331, USA}

\address[label2]{Department of Astronomy University of Maryland 1113
  Physical Sciences Complex College Park, Md 20742-2421}

\address[label3]{Instituto de Astronom\'i­a, Universidad Nacional
  Aut\'onoma de M\'exico, Apdo. postal 70-264 Ciudad Universitaria, D.F.,
  Mexico}

\begin{abstract}
  Gamma-ray bursts are a complex, non-linear system that evolves very
  rapidly through stages of vastly different conditions. They evolve
  from scales of few hundred kilometers where they are very dense and
  hot to cold and tenuous on scales of parsecs. As such, our
  understanding of such a phenomenon can truly increase by combining
  theoretical and numerical studies adopting different numerical
  techniques to face different problems and deal with diverse
  conditions. In this review, we will describe the tremendous
  advancement in our comprehension of the bursts phenomenology through
  numerical modeling. Though we will discuss studies mainly based on
  jet dynamics across the progenitor star and the interstellar medium,
  we will also touch upon other problems such as the jet launching,
  its acceleration, and the radiation mechanisms. Finally, we will
  describe how combining numerical results with observations from
  Swift and other instruments resulted in true understanding of the
  bursts phenomenon and the challenges still lying ahead.
\end{abstract}


\end{frontmatter}

% \linenumbers

\section{Introduction}
\label{S:intro}

Numerical simulations have played a major role in the understanding of
gamma-ray burst (GRB) studies in the past decade. Even though it is
difficult to find a precise moment in which it all begun, the growing
evidence of association between long-duration GRBs and
core-collapse supernovae in the late 1990s \citep{woos93, gala98,
  pacz98, bloo99} arguably played a major role in supporting the need
for theoretical tools that could go beyond the approximations of
spherical symmetry and/or top-hat jets. Numerical simulations are now
used as a major tool in many aspects of the GRB phenomenology.

First, numerical methods are used to understand the properties of the
progenitor. Binary compact mergers are heavily studied as short GRB
progenitors \citep{ross07,giac12, giac13,ross13} and massive, fastly
spinning stars and their core-collapse are investigated as potential
long GRB progenitors \citep{woos06,yoon06,woos12}.  Numerical
simulations are also used to understand the jet launching from a
compact object, either a black hole or a magnetar
\citep{mcki07a,mcki07b,hari09,komi09,mcki09,naga09,hari10,tayl11,jani13,mcki13}. Subsequently,
numerical simulations are used to model the dynamics of both
magnetized \citep{bucc09,bucc12} and unmagnetized jets
\citep{macf99,aloy00,zhan03,zhan04,mizu06,mors07,mizu09,mors10,lope13,mizu13}. Numerical
simulations are finally used to model the prompt emission phase
\citep{peer06,lazz09,lazz10,lazz11,mizu11,vurm11,lazz13,lund13,lope14,lund14,chho15}
and, eventually, the afterglow
\citep{vane11,deco12a,deco12b,vane12,vane13}.

In this review we will concentrate on the hydrodynamical aspect of
simulations, focusing on the interaction between the jet and the
progenitor star and its consequences on the jet dynamics, propagation,
and radiation mechanism. We remind the reader to the above references
for a more complete discussion of the various numerical techniques and
physical problems addressed.

\section{Inside the star: ploughing through}

\begin{figure}
\centerline{
\includegraphics[width=0.8\textwidth]{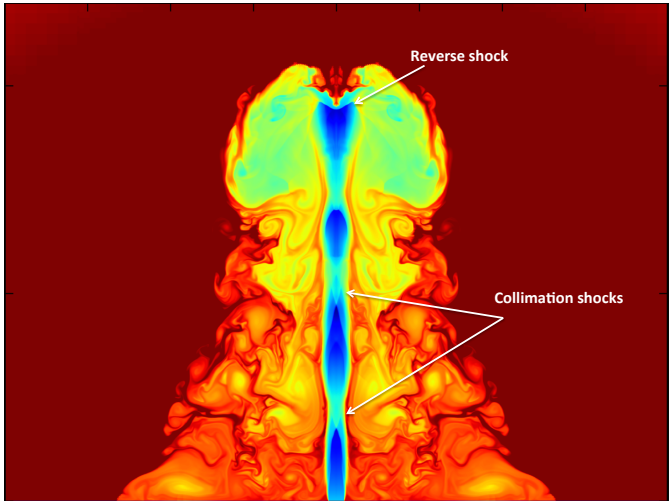}}
\caption{{False-color rendering of a relativistic jet expanding in the
  core of a massive star. Red colors show high-density while blue
  colors show low-density regions. The reverse shock that decelerates
  the jet material and the tangential collimation shocks are
  indicated. The forward bow-shock propagating in the interstellar
  materia is not shown. Adapted from \citep{lazz12}.}
\label{fig:jet_deep}}
\end{figure}

Hydrodynamical (HD) simulations of relativistic jets inside massive
stars have played a major role in our understanding of the GRB
phenomenology. They are based on the assumption that somehow the central
engine - being a black hole or a magnetar - is capable of producing a
jet with the adequate luminosity and entropy. The jet has to propagate
through a star that is mostly unchanged since core-collapse, its
free-fall time being longer than the typical GRB duration at radii
beyond $\sim10^9$~cm from the star's center. More controversial is the
jet composition at the jet's base, i.e. the inner boundary of the
simulation. Most simulations are HD and ignore the presence of magnetic
fields. This is a good approximation as long as the magnetization is
low. Since most jet launching mechanisms are heavily based on strong
magnetization, such an assumption has unclear validity. Simulating
unmagnetized jets, on the other hand, makes it possible to satisfy the
requirement of very high resolution in the boundaries between the
relativistic outflow and the surrounding star, a resolution that can be
achieved only with adaptive mesh techniques.

The first issue numerical simulations have to address is the propagation
of the jet inside the star. A known result is that the jet cannot expand
conically and accelerate proportionally to the radius inside the
progenitor star \citep{matz03}. Early GRB simulations
\citep{macf99,aloy00} showed that the jet head propagates
trans-relativistically, at few tens of per cent of the speed of light.
This speed depends very weakly on the jet and star properties and a
value $\beta_h=0.25$ for the jet-head speed gives an accurate
prescription for the propagation inside the star \citep{lazz12}. A
sub-luminal propagation speed also ensures that the jet is causally
connected with the star and the star material that accumulates in front
of the jet can move aside. Numerical results can be qualitatively
reproduced by analytical models \citep{mors07,brom07,brom11}. Even the
most advanced analytical models, however, assume cylindrical symmetry
and do not include important effects such as vortex shedding, multiple
tangential shocks, and turbulence. As a consequence, they cannot exactly
reproduce some simulations detail and fail to precisely predict even the
jet head expansion velocity inside the progenitor star \citep{lazz12}.

\begin{figure}
\centerline{
\includegraphics[width=0.8\textwidth]{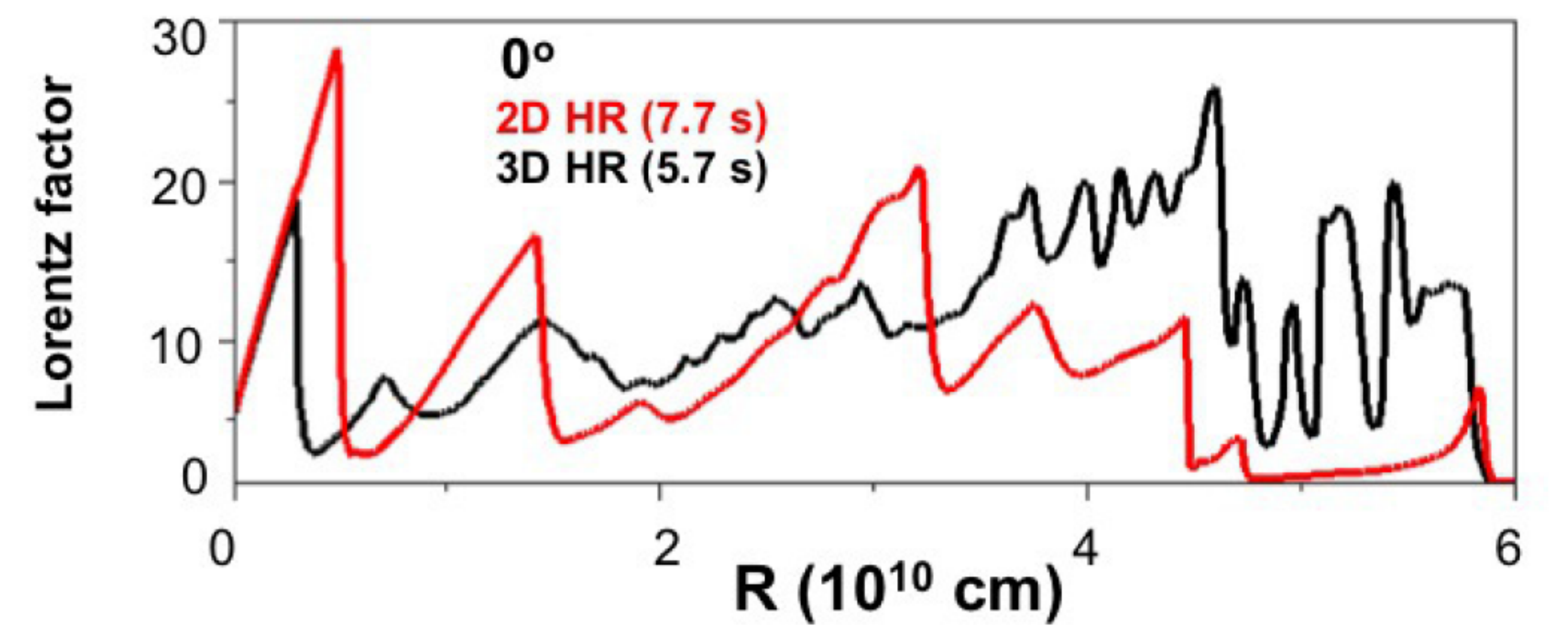}}
\caption{{Radial profile of the Lorentz factor of jets propagating in
    massive stars at the time of their breakout off the star's
    surface. Results from a 2D simulation (red) and a 3D simulation
    (Black) are compared, showing how 3D produces a more complex
    profile due to the presence of multiple minor shocks rather than a
    few strong ones. Adapted from \citep{lope13}.}
\label{fig:gamma}}
\end{figure}

One important consequence of a relatively slow jet propagation inside
the star is the creation of a cocoon that surrounds the jet. An amount
of energy
\begin{equation}
  E_{\mathrm{Cocoon}}=L_j\left(t_{\mathrm{bo}}-\frac{R_\star}{c}\right)\sim
  L_j\frac{R_\star}{c\beta_h}=10^{52}L_{j,51}R_{\star,11}\,\,\,\mathrm{erg}
\end{equation}
is deposited in the cocoon and, from the cocoon, is transferred to the
star. $L_j$ is the engine luminosity, $t_{\mathrm{bo}}$ is the jet
breakout time, $R_\star$ is the progenitor star's radius, and $\beta_h$
is the jet's head propagation speed in units of the speed of light.
$L_{j,51}$ and $R_{\star,11}$ are the luminosity and stellar radius
normalized by 10$^{51}$~erg~s$^{-1}$ and 10$^{11}$~cm, respectively.
Note that once the jet head has broken out ont he star's surface, all
the jet behind the head does exit the star, accounting for the
$R_\star/c$ term in the equation above. The energy deposited in the
cocoon is therefore enough to unbind the stellar material. However,
because the jet deposits the energy in the star far from the core, the
explosion might be darker than a normal core-collapse supernova (CCSN).
This is due to the lack of newly synthesized $^{56}$Ni, whose decay
powers the light curve of ``normal'' CCSNe. The presence of jets,
however, changes the energy distribution in the ejecta, producing
explosions with fast ejecta that can explain bright radio emission in
some supernovae \citep{lazz12}.

\begin{figure}
\centerline{\includegraphics[width=0.8\textwidth]{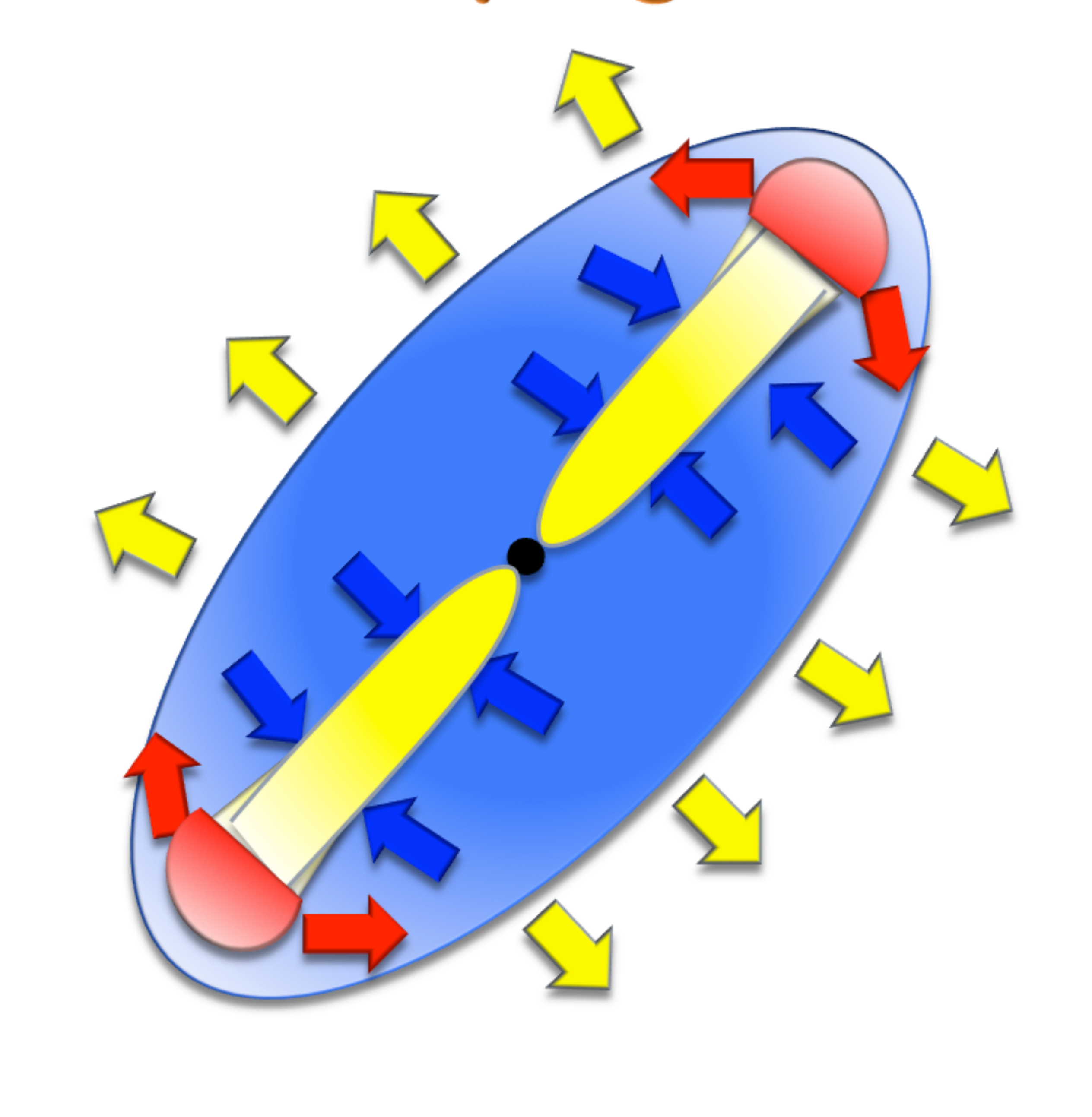}}
\caption{{Cartoon of the jet propagation dynamics inside a massive
    star. The jet (yellow) is launched. It pushes against the star
    material and a bow shock (Red) develops, shoving hot jet and star
    material to the side, feeding a uniform pressure cocoon
    (blue). The cocoon pressure pushes against the star, unbinding it,
  and the jet, collimating it into a smaller angle.}
\label{fig:cartoon}}
\end{figure}

A firm result of simulations, independent of the code and of the jet
and star properties, is the complexity of the jet profile. The jet is
characterized by the presence of multiple shocks
(Figure~\ref{fig:jet_deep}). There is a reverse shock that decelerates
the expanding jet as a consequence of the bow shock at the jet's
head. There are, however, several collimation shocks behind the
reverse shock as well. These are tangential shocks that are produced
by the interaction of the jet with the cocoon. As a consequence of the
presence of collimation shocks the jet's Lorentz factor is not uniform
behind the reverse shock, but it has a characteristic sawtooth shape
(Figure~\ref{fig:gamma}). A cartoon showing the various components of
the jet-star interaction dynamics is shown in
Figure~\ref{fig:cartoon}.

\begin{figure}
\centerline{
\includegraphics[width=0.8\textwidth]{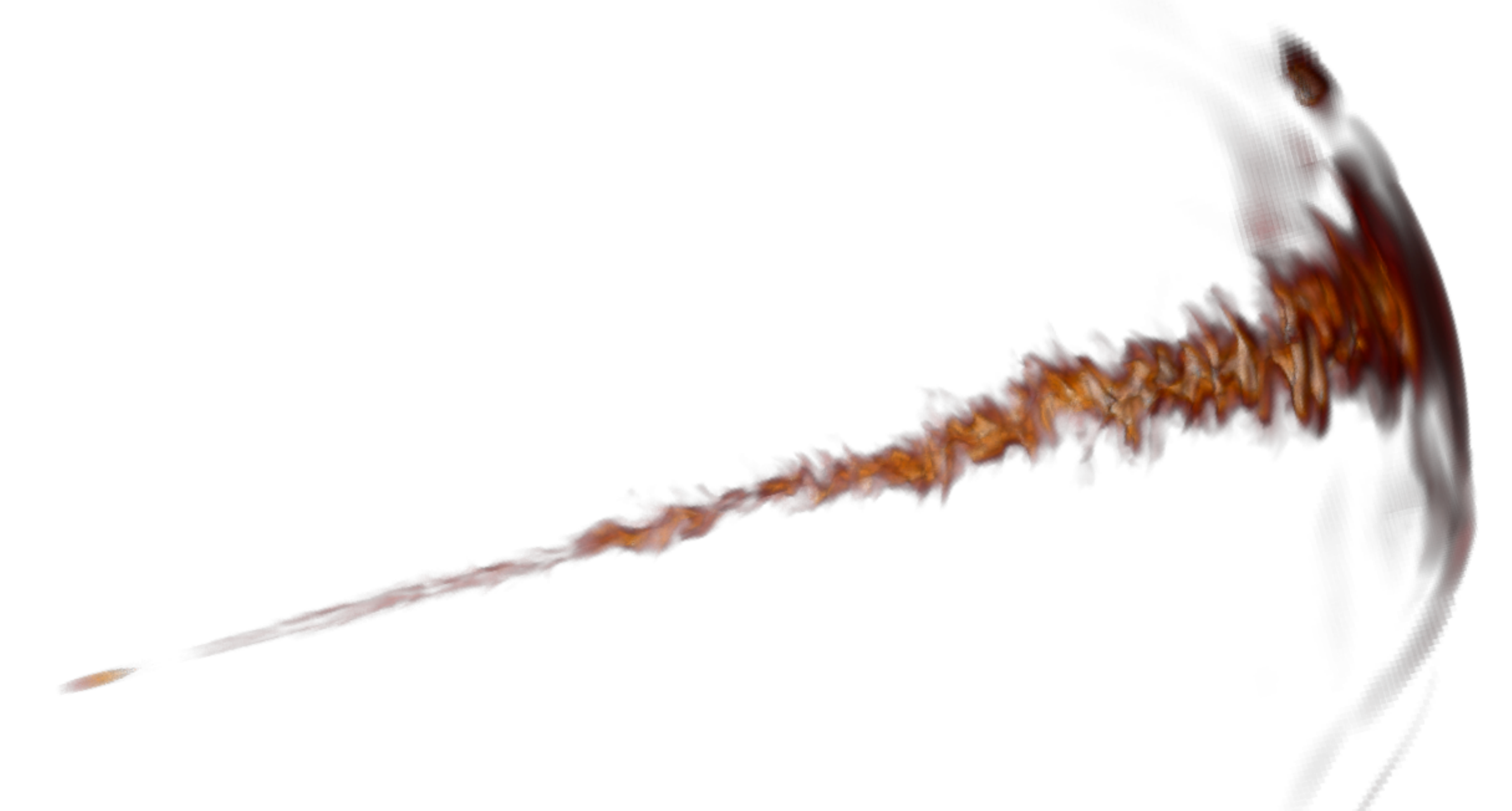}}
\caption{{Volume rendering of the velocity of a long GRB jet as it
    erupts off the surface of the progenitor star.}
\label{fig:3D}}
\end{figure}

Initial simulations of the jet propagation were performed in
cylindrical symmetry in two dimensions
\cite{macf99,aloy00,mors07}. More recently, full 3D simulations have
become possible. They show interesting features and more complexity in
the jet dynamics. One important limitation of 2D simulations is the
``plug instability'', an effect whereby any overdensity of ambient
medium that accumulates ahead of the jet next to the axis is trapped
and creates an obstacle. As a consequence the system develops two
plumes of low-density, high-temperature material at large polar angles
(see, e.g., Figure~1 in \citet{lazz10b}). This instability is seen in
jets from both constant and variable engines \citep{lope14}. 3D
simulations have shown that the jet, instead, travel through the path
of least resistance, its head moving round the polar axis to avoid
over-densities in the progenitor star or induced by the bow shock
itself \citep{zhan04,lope13}. As a consequence, the collimation shocks
are also reduced in size and intensity, producing a more complex
structure and a smoother profile of the Lorentz factor
(Figures~\ref{fig:gamma},~\ref{fig:3D}).

\section{Outside the star: free expansion...almost}

A second important stage of a GRB jet is its expansion once it has
left the progenitor star. The jet is expected to be freely expanding
at this point, accelerating proportionally to its radius until reaching
the asymptotic Lorentz factor of several hundreds (e.g.,
\citep{pira99}). However, simulations have shown that the interactions
with the stellar material carries on after the jet has left the
star. External material is provided by the expanding cocoon that
propagates out of the star at close to the speed of light but has a
significantly higher density than the jet. This causes the jet to
accelerate at a significantly slower rate than a free-expanding jet
and ensures the survival of tangential collimation shocks well beyond
the stellar surface \cite{lazz09}. It has also been shown that the
interaction of the star with the jet can imprint new variability
features on the jet, especially for long lived engines.

\begin{figure}
\centerline{
\includegraphics[width=0.4\textwidth]{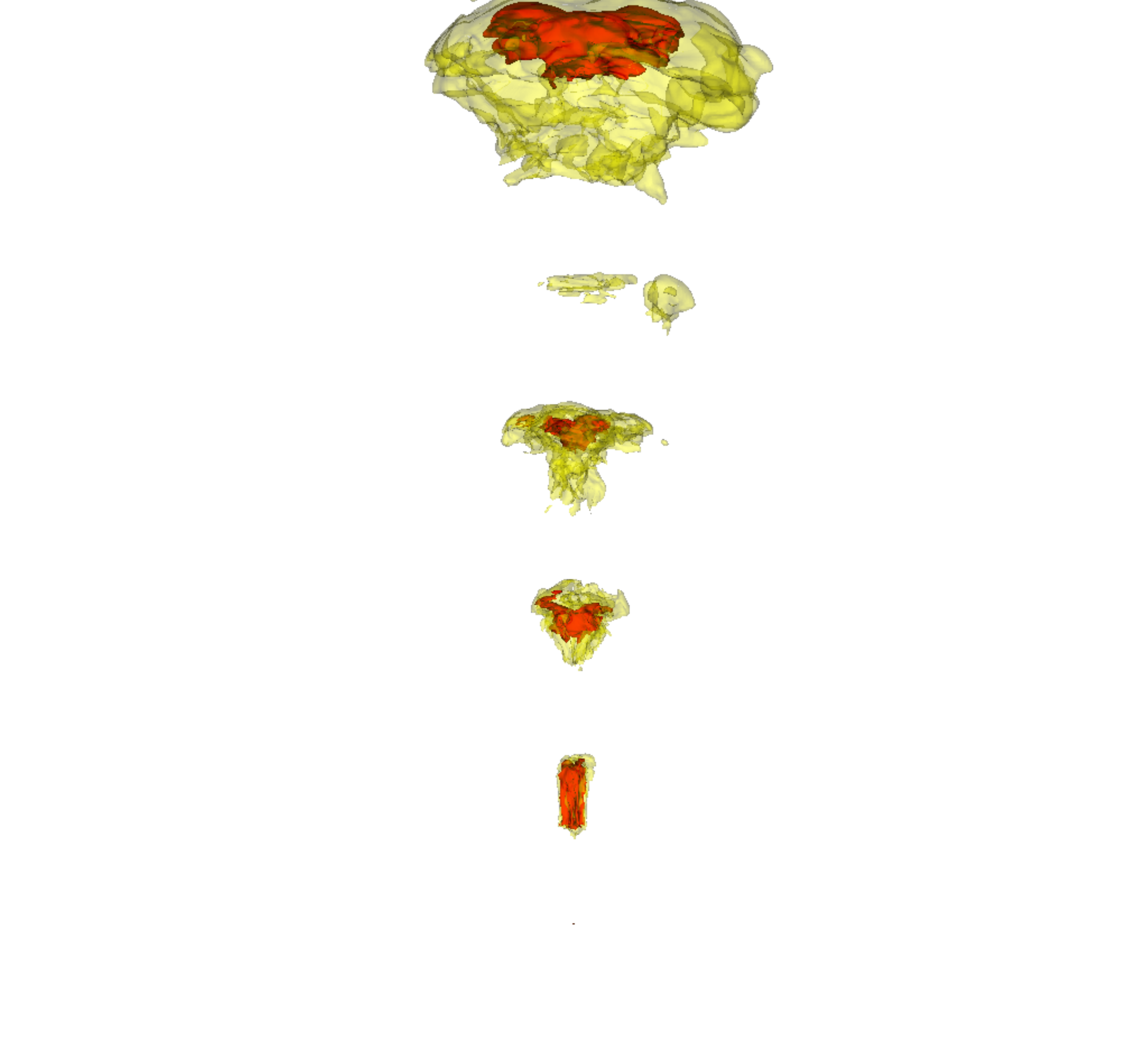}
\includegraphics[width=0.4\textwidth]{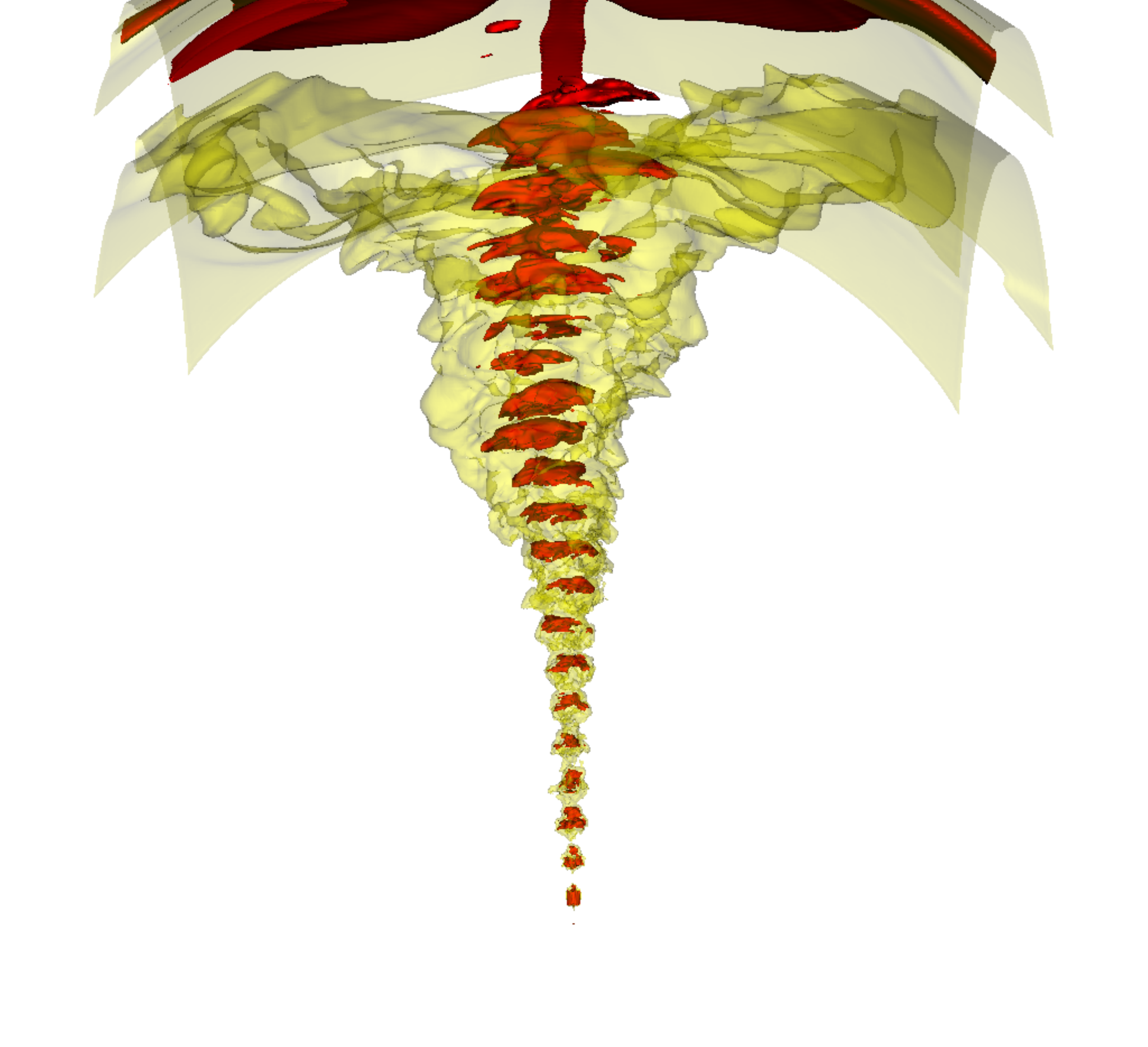}
                }
\caption{{Comparison between the evolution of jets from engines with
    different dead times. Red colors show relativistic expansion,
    while yellow colors show mildly relativistic expansion. The jet on
    the left is produced by an engine with activity and dead times of
    0.5 seconds. The jet on the right is produced by an engine with
    activity and dead times of 0.1 seconds.}
\label{fig:onoff}}
\end{figure}

Simulations of variable GRB jets have been performed in 2D
\citep{mors10, lope14} and, more recently in 3D (Lazzati et al. in
preparation). They show that the jet star interaction can work in
different ways, depending on the time-scale of the variability.
\begin{itemize}
\item If the engine is characterized by very long time-scale
  variability, longer than a few seconds, the interaction with the
  star does not affect the long-term variability, but adds a
  few--second timescale to the jet energy outflow \citep{mors10}.
\item If the engine has very high frequency variability (faster than a
  few tens of Hz), this is left almost unmodified by the jet
  propagation through the progenitor and can be translated, almost
  pulse to pulse, in the jet luminosity profile \citep{mors10}.
\item If the engine has variability on time-scales of seconds, they
  interact destructively with the progenitor star variability
  timescal. For a variable jet that is able to break out of the star,
  the first active periods are destroyed as they create the funnel
  through the stellar progenitor \citep{lope14}.  A set of 3D
  simulations was performed to investigate the role of jet variability
  in the stability of jet propagation. It showed that in the cases in
  which the dead times of the jet are of the order of $\sim1$ seconds,
  the star has time to close the funnel previously dug by the jet. As
  a consequence, energy has to be spent again in reopening the channel
  and the duration of activity times, as seen by a distant observer,
  is reduced (Lazzati et al. in preparation). A comparison between the
  jet of an engine with 0.5~s dead times and the jet of an engine with
  0.1~s dead times is shown in Figure~\ref{fig:onoff}. while the jet
  from the engine with faster variability preserves the duration of
  the active and dead times (equal spacing and durations of red and
  yellow phases), the jet from an engine with longer dead times needs
  to use the energy of the active pulses to re-open the funnel. As a
  consequence active pulses reach the star surface with significantly
  reduced duration producing an asymmetry between active and dead
  times in the jet luminosity at large radii. Such an asymmetry is
  imprinted in the light curve as well, and could explain the
  detection of longer than usual dead times in some long-duration
  BATSE GRBs \citep{naka02}.
\end{itemize}

\begin{figure}[!t]
\centerline{
\includegraphics[width=\textwidth]{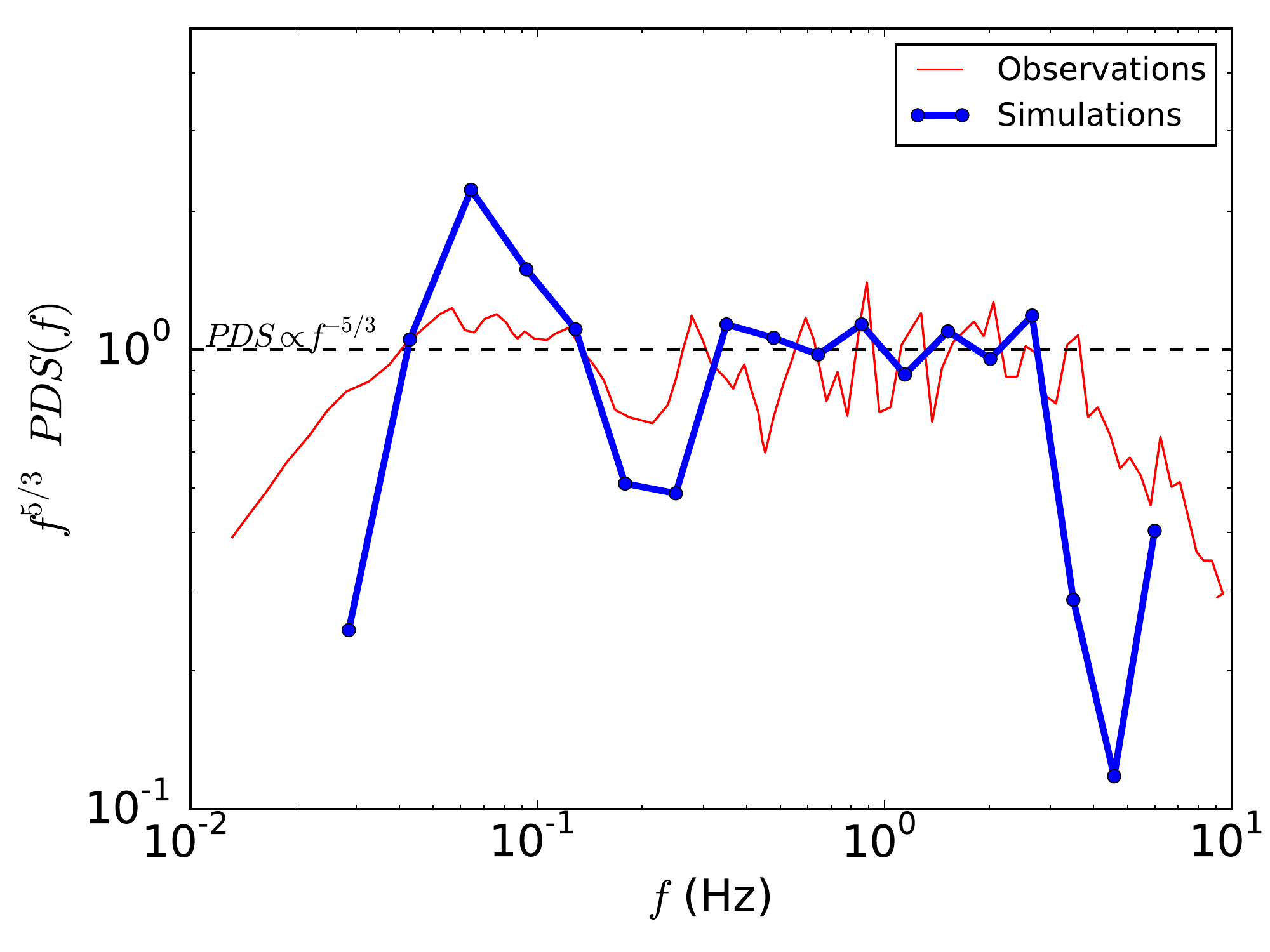}}
\caption{{Comparison between the average power density spectrum of a
    sample of BATSE GRBs \citep{belo98} and the average power density
    spectrum of synthetic light curves from hydrodynamic simulation of
  GRB jets \citep{mors10}.}
\label{fig:pds}}
\end{figure}

A significant result of these simulations is the fact that the resulting
light curves have variability properties analogous to the ones of
observed GRBs \cite{mors10, lope14}. Figure~\ref{fig:pds} shows a
comparison between the power spectrum of light curves from 2D HD
simulations and those of observed BATSE long-duration GRBs. Simulations
reproduce the average slope ($\propto\nu^{-5/3}$), the low frequency
cut-off and the high-energy cutoff seen in observations. Especially the
low-frequency cut-off, entirely set by the interaction with the
progenitor star, is significative of the role of the star in shaping the
light curve of bursts.

Finally, the jet star interaction might be responsible for X-ray
flares in the light curve of burst afterglows \citep{chin07}. Any
instability in the jet pressure against the star is amplified by the
star's response, leading to a time-duration correlation consistent
with observations \citep{lazz11b}.

\section{The radiative stage and ensamble correlations}

The radiation mechanism of GRBs is perhaps still the most
controversial aspect of the whole phenomenon. The standard view of a
synchrotron-dominated spectrum \citep{pira99} is giving way to a more
elaborated scenario in which advected radiation released at the
photosphere dominates or, at least, contributes substantially to the
light curve energy budget \citep{peer06,guir11,lazz13}. Within this
scenario, meaningful light curves and spectra can be straightforwardly
calculated from the results of numerical simulations
\citep{lazz09,mizu11,cues15a,cues15b}. The underlying assumption is
that radiation and matter are in equilibrium until some radius in the
sub-photospheric region where they decouple and evolve independently
thereafter. Assuming also that the radiation has a thermal spectrum,
the bolometric luminosity and peak frequency of the emission can be
extracted from local HD quantities (energy density) and
boosted to the observer frame given the local velocity.

\begin{figure}[!t]
\includegraphics[width=0.48\textwidth]{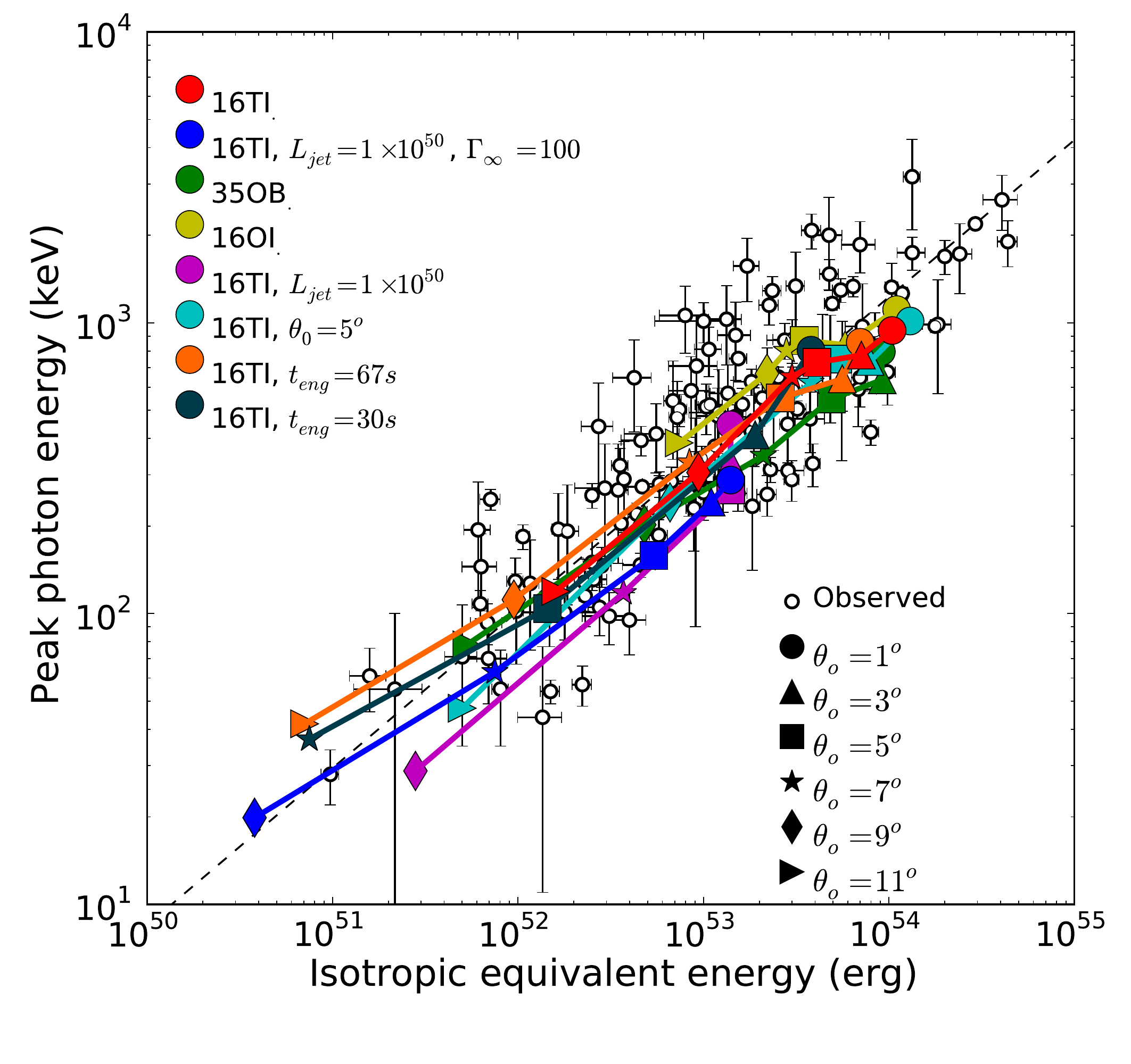}
\includegraphics[width=0.48\textwidth]{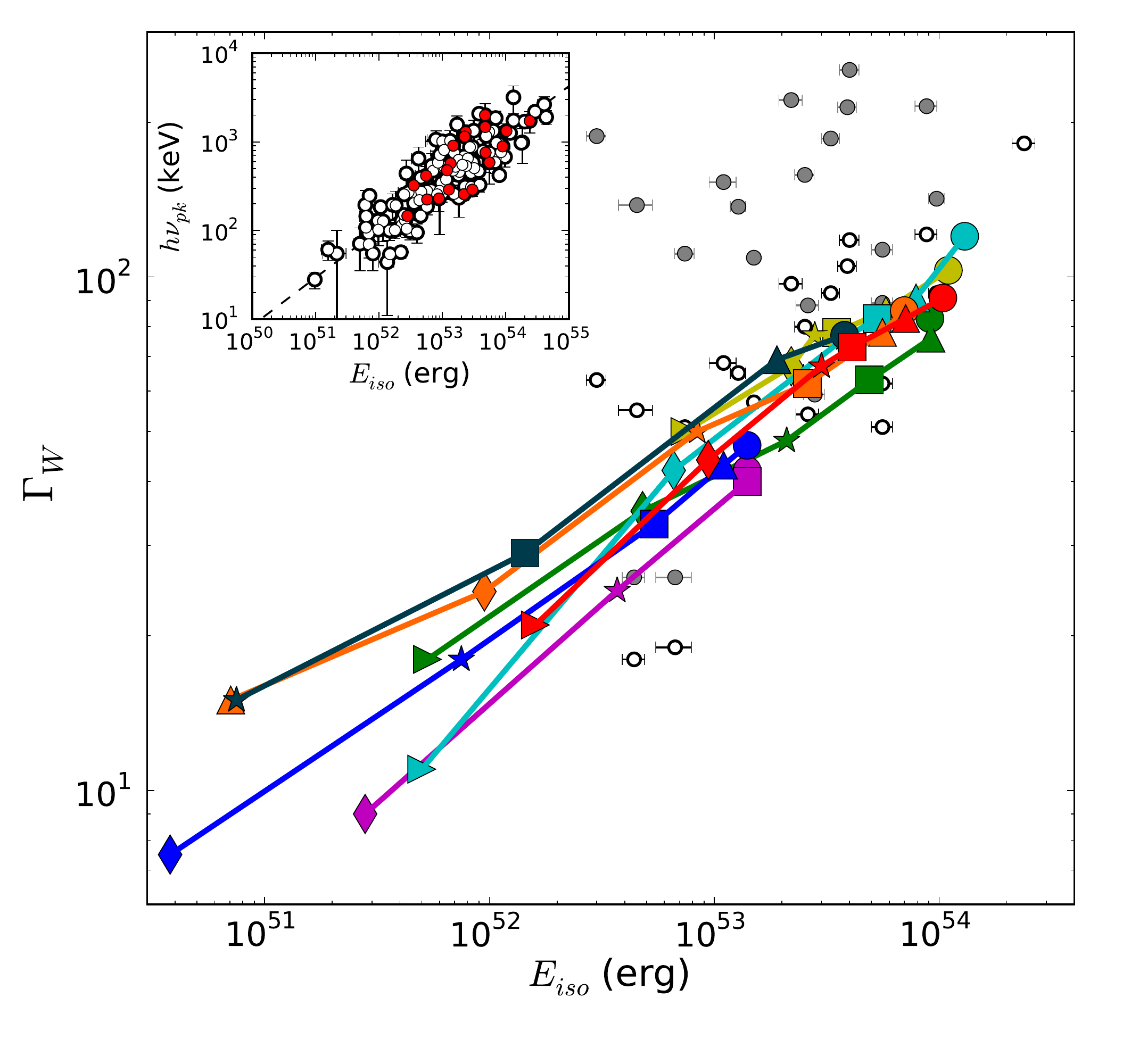}
\parbox{0.49\textwidth}{\includegraphics[width=0.48\textwidth]{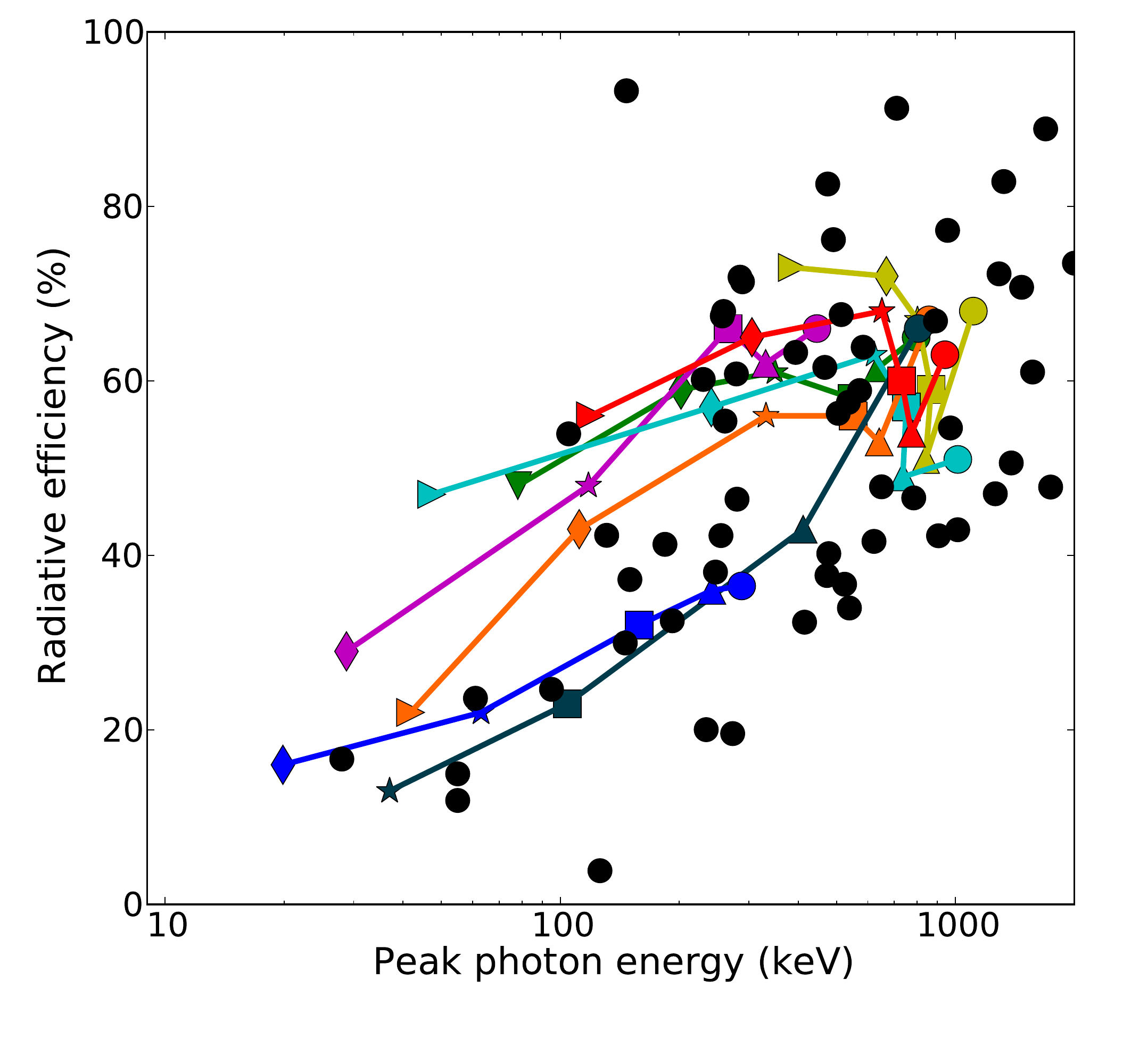}}
\parbox{0.49\textwidth}{
  \caption{{Observational ensamble correlations (black symbols)
      compared to the results of numerical simulations. The top left
      panel shows the Amati correlation, the top right panel the
      Lorentz factor-energy correlation and the bottom left panel
      shows the efficiency peak frequency correlation. Adapted from
      \citep{lazz13}.}
\label{fig:amati}}}
\end{figure}

\begin{figure}[!t]
\centerline{
\includegraphics[width=0.8\textwidth]{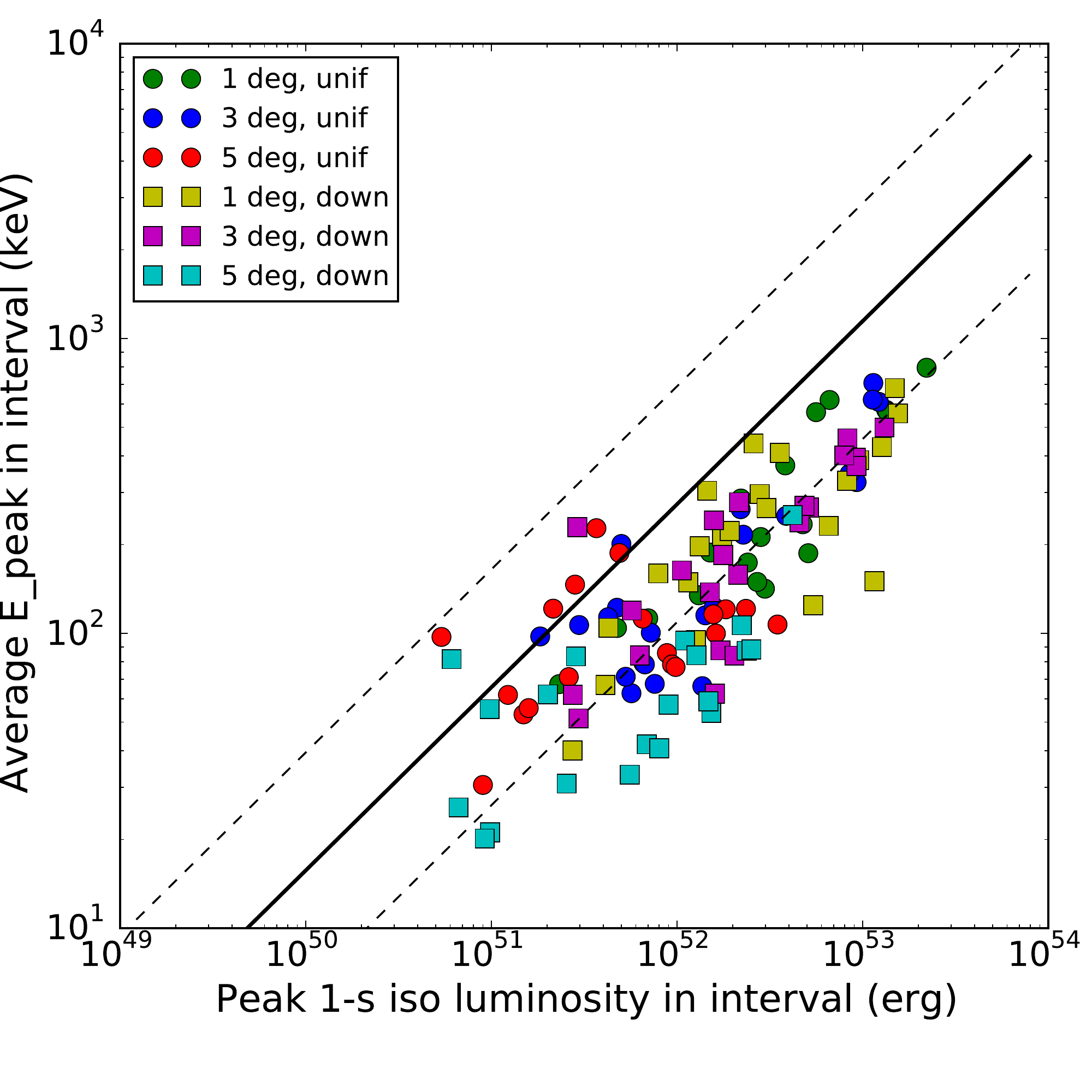}}
\caption{{Golenetskii correlation. Results from numerical simulations
    (colored symbols) are overlaid on the best-fit observational
    correlation (solid line) and its 2-sigma uncertainty region
    (dashed lines). Adapted from \cite{lope14}.}
\label{fig:gole}}
\end{figure}

Such simulations have been very successful in reproducing ensamble
results of GRB emission based on a very few assumptions on the
properties of the progenitor stars and their central engines. The
Amati correlation \citep{amat02} was shown to be reproduced as an
effect of the viewing angle, robust to changes of jet and progenitor
star properties\citep{lazz09,lazz13}. The same set of simulations
reproduced the Lorentz factor-isotropic energy correlation
\citep{lian10,ghir12} and predicted the existence of a correlation
between radiation efficiency and peak frequency, later found in Swift
data. The three observational correlations are shown in
Figure~\ref{fig:amati} with the theoretical data overlaid.  Another
correlation that can be explained from simulations within the
photospheric scenario is the Golenetskii correlation \citep{lope14},
the correlation between the peak energy and the luminosity in finite
intervals of GRB light curves \citep{gole83,ghir10,lu12}. This
correlation is particularly significant given the debate over the
reliability of the Amati correlation. A set of simulations with
intermittent engines showed that the general behavior of the
Golenetskii correlation could also be produced, albeit with a marginal
offset in normalization (Figure~\ref{fig:gole}).

\section{Spectral calculations}

Despite all the success seen above, radiation codes coupled to
numerical simulations cannot relax the assumption of thermal
equilibrium and cannot therefore reproduce the broad-band non-thermal
character of observed GRB spectra. Detailed spectral models come
therefore from one-zone simulations in which the diversity of the
dynamic conditions seen in HD simulations is neglected in
favor of a more detailed spectral modeling. In some cases, one
dimension of complexity is maintained, assuming a radial evolution of
the jet in free expansion \citep{peer06,gian06}. In other cases, more
sophisticated radiation treatment can be applied only to a single zone
\citep{chho15}.

\begin{figure}[!t]
\centerline{
\includegraphics[width=0.8\textwidth]{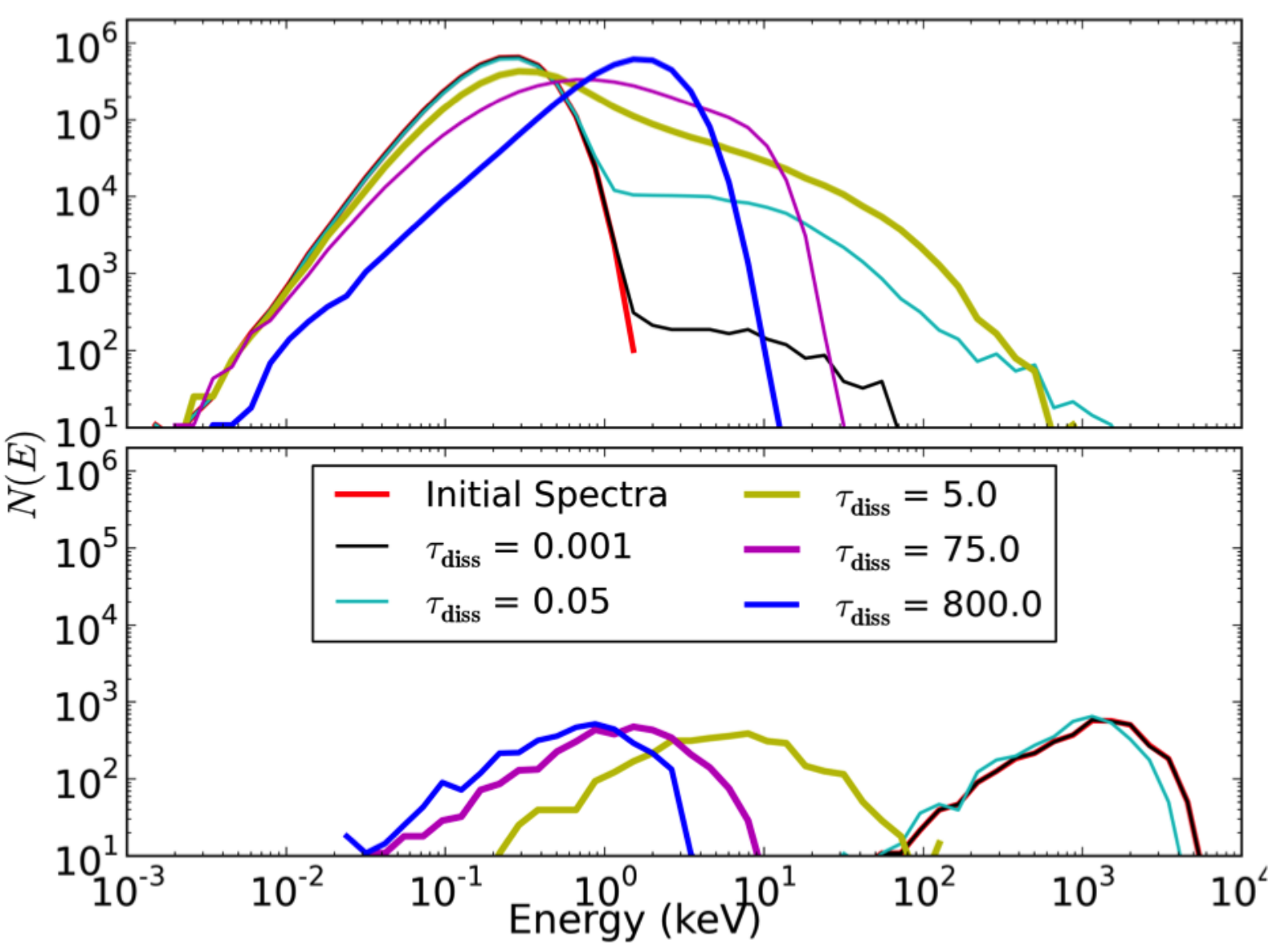}}
\caption{{Monte Carlo simulations of radiation-lepton relaxation
    following the sudden energization of the leptons by a dissipation
    event in a scattering dominated medium.}
\label{fig:MC}}
\end{figure}

Figure~\ref{fig:MC} shows the results of Monte Carlo simulations of
the radiation-matter relaxation following an event of sudden energy
dissipation in the lepton population. The relaxation is assumed to be
dominated by Compton and inverse Compton scattering and by pair
creation and annihilation. Calculations like these are necessary since
GRB jets at the photosphere are radiation dominated and one cannot
assume that the electron have a well defined temperature. As seen in
Figure~\ref{fig:MC}, both the photon and the electron energy
distributions have a markedly non-thermal character in the transition
period while equilibrium is restored. Since the scattering take place
in a relativistically expanding medium, in which photons and electrons
propagate at very small angle, the scattering rate is greatly
inhibited and the non-thermal phase can last for a significant amount
of time.

\section{Summary and conclusions}

In summary, numerical simulations of GRB outflows have allowed us to
gain a much deeper insight in the phenomenon. Comparisons with large
sample of data such as the BATSE and Swift catalogs have been pivotal
in refining the parameters and physics that is needed to explain GRB
observations. In this review we concentrated on the HD aspect of the
GRB phenomenon and briefly touched on the prompt radiation
phase. Other phases, progenitor, central engine, and afterglow have
also benefited from numerical studies and we remind the readers to the
extensive literature cited in the Introduction for a complete review
of numerical studies of GRBs.

\section{Acknowledgements} This research was partially supported by the
Fermi GI program grant NNX12AO74G and Swift GI program grant NNX13A095G
(DL and DLC). BJM is supported by an NSF Astronomy and Astrophysics
Postdoctoral Fellowship under award AST-1102796.

\bibliographystyle{elsarticle-harv}



\end{document}